\documentclass[aps,prl,reprint,superscriptaddress]{revtex4-1}

\usepackage{graphicx}
\usepackage{amssymb}
\usepackage{amsmath}
\usepackage{epstopdf}

\newcommand{\abs}[1]{\left|#1\right|}
\newcommand{\K}{\mathrm K}

\newcommand{\mV}{\mathrm{mV}}
\newcommand{\meV}{\mathrm{meV}}

\newcommand{\kBT}{k_{\mathrm B}T}
\newcommand{\kohm}{\mathrm k\Omega}

\newcommand{\gzero}{g_{\mathrm{0}}}

\newcommand{\MHz}{\mathrm{MHz}}

\newcommand{\GHz}{\mathrm{GHz}}

\newcommand{\fr}{\omega_{\mathrm{r}}}
\newcommand{\fd}{\omega_{\mathrm{d}}}

\newcommand{\VLG}{V_{\mathrm {LG}}}
\newcommand{\VMG}{V_{\mathrm {MG}}}
\newcommand{\VRG}{V_{\mathrm {RG}}}

\newcommand{\Vsd}{V_{\mathrm {SD}}}
\newcommand{\tcoupl}{t_{\mathrm c}}
\newcommand{\zo}{Z_{\mathrm 0}}
\newcommand{\Zm}{\ensuremath{Z_\mathrm{Match}}}
\newcommand{\Gm}{\ensuremath{G_\mathrm{Match}}}

\newcommand{\Gcnt}{\ensuremath{G_\mathrm{CNT}}}
\newcommand{\GStub}{\ensuremath{G_\mathrm{Loss}}}
\newcommand{\Geff}{\ensuremath{G_\mathrm{eff}}}

\begin{document}

\title{Clean carbon nanotubes coupled to superconducting impedance-matching circuits}
\author{V.~Ranjan}
\affiliation{Department of Physics, University of Basel, Klingelbergstrasse 82, 4056 Basel, Switzerland}
\author{G.~Puebla-Hellmann}
\affiliation{Department of Physics, University of Basel, Klingelbergstrasse 82, 4056 Basel, Switzerland}
\affiliation{Department of Physics, ETH Zurich, Otto-Stern-Weg 1, 8093 Zurich, Switzerland}
\author{M.~Jung}
\author{T.~Hasler}
\author{A.~Nunnenkamp}
\affiliation{Department of Physics, University of Basel, Klingelbergstrasse 82, 4056 Basel, Switzerland}
\author{M.~Muoth}
\affiliation{Department of Mechanical and Process Engineering, ETH Zurich, Tannenstrasse 3, 8093 Zurich, Switzerland}
\author{C.~Hierold}
\affiliation{Department of Mechanical and Process Engineering, ETH Zurich, Tannenstrasse 3, 8093 Zurich, Switzerland}
\author{A.~Wallraff}
\affiliation{Department of Physics, ETH Zurich, Otto-Stern-Weg 1, 8093 Zurich, Switzerland}
\author{C.~Sch\"{o}nenberger}
\email{Christian.Schoenenberger@unibas.ch}
\affiliation{Department of Physics, University of Basel, Klingelbergstrasse 82, 4056 Basel, Switzerland}

\date{\today}

\begin{abstract}

\textbf{Coupling carbon nanotube devices to microwave circuits offers a significant increase in bandwidth and signal-to-noise ratio.
These facilitate fast non-invasive readouts  important for quantum information processing, shot noise and correlation measurements.
However, creation of a device that unites a low-disorder nanotube with a low-loss microwave resonator has so far remained a challenge,
due to fabrication incompatibility of one with the other. Employing a mechanical transfer method, we successfully couple a
nanotube to a gigahertz superconducting matching circuit and thereby retain pristine transport characteristics such as the control over
formation of, and coupling strengths between, the quantum dots. Resonance response to changes in conductance and susceptance further
enables quantitative parameter extraction. The achieved near matching is a step forward
promising high-bandwidth noise correlation measurements on high impedance devices such as quantum dot circuits.}

\end{abstract}

\maketitle

Artificial two-level systems such as Josephson junction qubits coupled to superconducting microwave cavities~\cite{Wallraff2004,Clarke2008} have allowed for unprecedented break-throughs in quantum information processing. Inspired by these results, other solid state systems have been coupled to resonant circuits, such as double quantum dots in semiconductors~\cite{Frey2012,Toida2013, Petersson2012, Deng2013, Viennot2014} as potential qubits, and molecules or impurity spin ensembles~\cite{Rabl2006,Kubo2010,Schuster2010,Wu2010} as quantum memories. In particular, carbon nanotubes (CNT) have recently demonstrated their potential as low disorder one-dimensional electron systems~\cite{Pei2012,Waissman2013,Jung2013}, which have been used to probe the physics of spin-orbit~\cite{Kuemmeth2008} and electron-phonon coupling~\cite{Benyamini2014} as well as to perform
initialization and manipulation of spin qubits~\cite{Laird2013}. Carbon nanotubes suspended over local gates not only offer a decoupling from the surface but also mechanical resonances with high quality factors~\cite{Huttel2009}, creation of arbitrary local potentials and tight confinement with charging energies in excess of 50 meV~\cite{Jung2013}.

Nevertheless, coupling CNTs to half-wave resonators~\cite{Delbecq2011, Delbecq2013,Viennot2014} and other types of $LC$ circuits~\cite{Roschier2004,Tang2007,Lechner2010,Chorley2012} faces a significant challenge: achieving a low microwave loss device while preserving the ideal transport characteristics of pristine CNTs in a geometry which allows full control over charge confinement. While CNT devices with clean transport spectra can be obtained using a growth-last approach~\cite{Cao2005}, the low yield and high temperatures ($\sim 900^\circ\mathrm{C}$) involved prohibit the use of this method for fabricating superconductor-CNT hybrid devices. Recently, the transfer of CNTs from a growth substrate to a separate microwave device, followed by standard lithography has been demonstrated~\cite{Viennot2014a}, solving issues with material compatibility but not addressing limited yield and unclean transport spectra. As such, the combination of low loss microwave circuits with clean CNTs has not been achieved so far, impeding the way toward experiments involving the interplay of electrons in CNTs with microwave photons.

Here, we couple a locally tunable suspended CNT quantum device to an impedance-matching circuit based on superconducting transmission lines. Different to previous works~\cite{Delbecq2011, Delbecq2013,Viennot2014} where half wave resonators are employed for dispersive and a minimal invasive measurement, our circuit is aimed at providing an efficient channel to transfer (collect) microwave radiation into (from) a quantum device. Additionally, the circuit offers bandwidths (BW) in the MHz range even for device impedances on the
order of 1M$\Omega$. These features, on one hand, allow us to perform high BW measurements for deducing both conductance and susceptance changes in the quantum device at GHz frequencies, and on another hand, provide near unity collection of emitted radiation power for fast shot noise measurements. Through a mechanical transfer process~\cite{Muoth2013}, we place the CNT on the finished microwave device in the last step. Such a selective assembly technique~\cite{Waissman2013} allows us to address the mentioned fabrication and yield issues while obtaining clean transport spectra in combination with low microwave loss circuits. We employ local gates to demonstrate a high degree of control over the formation of double dots in the ambipolar regime and perform RF measurements of the device susceptibility. We are able to tune the inter-dot coupling strength and extract values on the order of $\GHz$ using the phase response of the microwave circuit. By performing simultaneous measurements of resistance and complex impedance, we observe good quantitative agreement between DC conductance and RF measurements.\\

%%%%%%%%%%%%%%%%%%%%%%%%%%%%%%%%%%%%%%%%%%%%%%%%%%%%%%%%%%%%
%Figure1
%%%%%%%%%%%%%%%%%%%%%%%%%%%%%%%%%%%%%%%%%%%%%%%%%%%%%%%%%%%%
\begin{figure}[t!]
\centering
\includegraphics{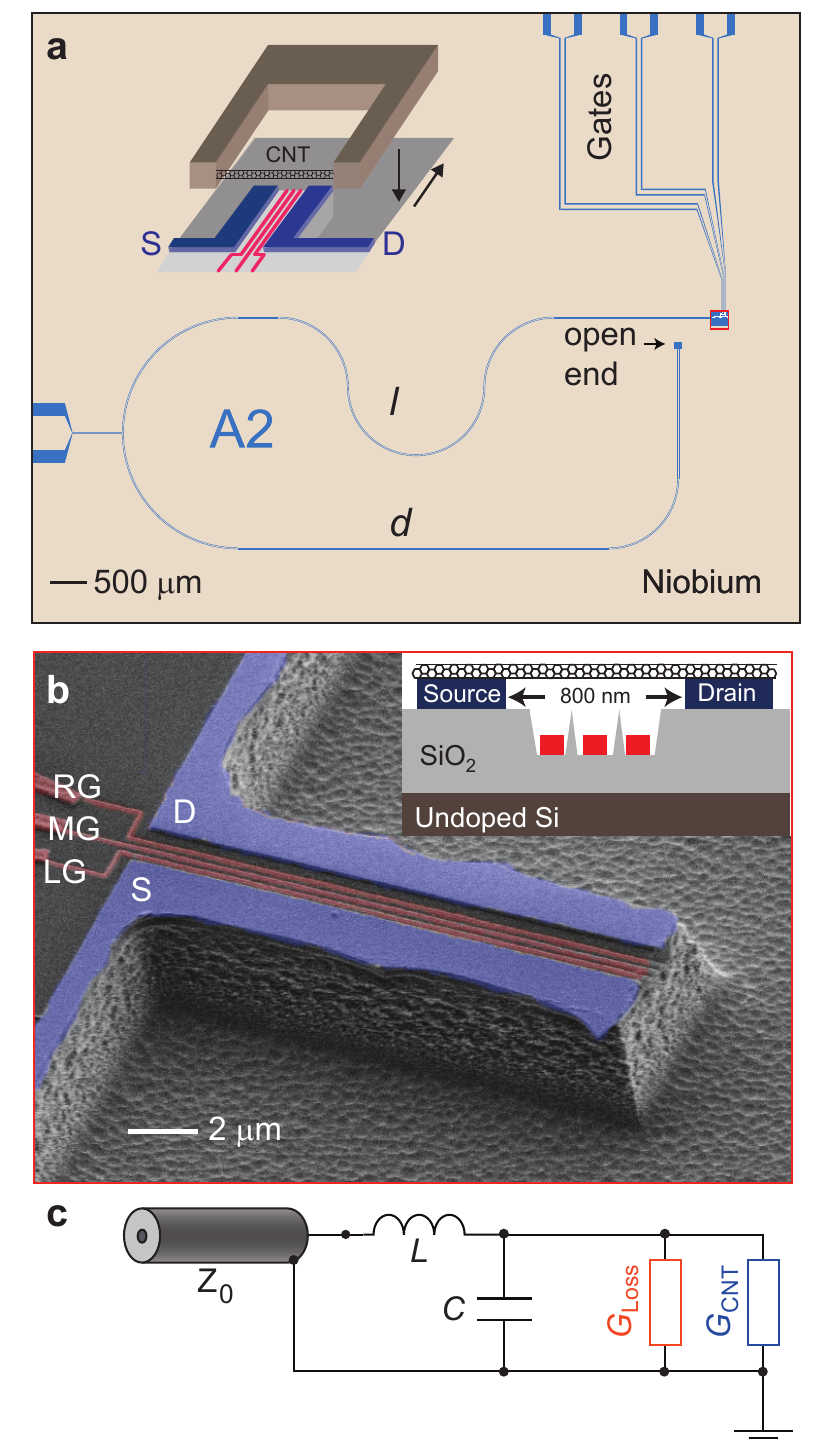} % for preprint
  \label{Figure1}
  \caption{\textbf{Device layout.} 
	(\textbf{a}) Design of an impedance-matching circuit based on two transmission
  lines with slightly different lengths $l$ and $d$, both close to $\lambda/4$. The line $d$ terminates in an open end
  while the central conductor of $l$ connects to the source (S) of the CNT device. A false-color SEM image is shown in
  (\textbf{b}) with left (LG), middle (MG) and right (RG) bottom gates. CNTs are transferred mechanically from fork structures
  to the bonded device chip in the final fabrication step, schematically shown in the inset of (a). The sketch displays relative
  position of CNTs with fabricated gates and palladium contacts. The drain (D) is connected to the ground plane of the Niobium
  film, not visible here. 
	(\textbf{c}) Equivalent matching circuit based on lumped elements. $L$ and $C$ are determined by the
  lengths $l$ and $d$, while the phenomenological conductance $\GStub$ captures the loss of the transmission lines. $\Gcnt$ is
  the device conductance of the CNT.}
\end{figure}

\noindent \textbf{Results}

\noindent \textbf{Mechanical transfer of CNT.} A key advance in this hybrid-CNT device is the implementation of a mechanical transfer at ambient conditions. The same marks the last step of sample fabrication. After the complete RF and DC circuitry are fabricated, the chip is bonded on a printed circuit board and cleaned with a weak argon plasma to remove oxides and fabrication residue from the contacts. We grow CNTs on the fork-like structures of a separate chip, which allow a resist- and ebeam radiation free mechanical transfer of the CNTs to the source/drain electrodes on a pillar structure (inset Fig.~1a). Forks are aligned with the pillar using an optical microscope of a micro-manipulator setup and successful transfers are monitored through voltage biased (200~mV) resistance measurements. We can then determine bandgap characteristics of transferred CNTs by measuring the conductance response to an applied gate voltage. Furthermore, we can remove tubes with unwanted characteristics, such as a metallic response, by applying a large bias voltage, and subsequently transfer and test another CNT on the same device. These capabilities make our method of transfer a deterministic one with unity device yield. Further details of the CNT growth and transfer method can be found in the Supplementary Note~2.\\  

%%%%%%%%%%%%%%%%%%%%%%%%%%%%%%%%%%%%%%%%%%%%%%%%%%%%%%%%%%%%
%Figure2
%%%%%%%%%%%%%%%%%%%%%%%%%%%%%%%%%%%%%%%%%%%%%%%%%%%%%%%%%%%%

\begin{figure}[b!]
\centering
\includegraphics{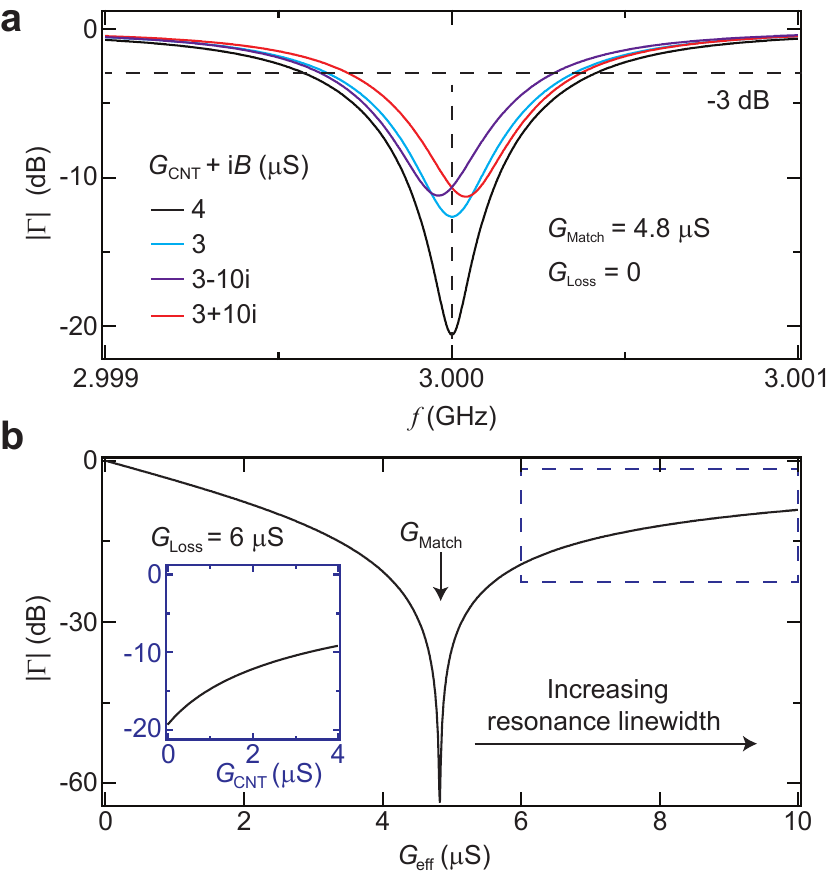} % for preprint
  \label{Figure2}
  \caption{\textbf{Calculated response of the matching circuit for different parameters.}
	(\textbf{a}) Reflection
    spectra of a loss-less stub tuner with 4.8 $\mu$S matched load at 3 GHz for different loads. Changing the real part
    of the load changes the linewidth and depth of the resonance, while changing the imaginary part mainly shifts the
    resonance frequency.
		(\textbf{b}) The reflectance at the resonance frequency as a function of a total effective load
    $(\Geff=\Gcnt+\GStub)$ shows a non-monotonic response and the smallest reflected signal at the matched load. Its
    characteristic changes to a monotonic response for $\GStub > \Gm$ as shown in the inset (same as the dashed box) for
    $\GStub=6~\mu \mathrm{S}$.}
\end{figure}
\noindent \textbf{Combining superconducting impedance matching circuits with gate controlled CNT devices.} A transmission line based impedance matching circuit, termed stub tuner~\cite{Pozar05}, can be realized by two transmission lines connected in parallel, with the device placed on one end, the other ending in an open circuit (Fig.~1a). Though simple lumped $LC$ circuits can also be employed, they demand a lengthy iterative design procedure when used at frequencies of GHz due to stray capacitances~$\sim$~fF. By utilizing transmission lines, we are not only able to minimize parasitic effects, such as stray capacitances, but are also able to compensate for them, as they only change the effective length of the lines.

The microwave response of the stub tuner is mainly determined by the lengths of the two lines $l$ and $d$, which are chosen such that a specific device impedance, called the matched load \Zm, will be transformed to the characteristic impedance of the line $\zo$ at one specific resonance frequency. At this load, the reflection coefficient is minimized,
implying maximum power transfer. To get the reflectance $\Gamma$, we calculate the impedance $Z_\mathrm{in}$ of the hybrid device, consisting of a parallel combination of the two transmission lines~\cite{Pozar05,Gabriel2012} as
\begin{equation}
Z_{\mathrm {in}} = Z_0\left( \tanh(\gamma d)+ \frac{Z_0+Z_{\mathrm {CNT}} \tanh(\gamma l) }{Z_{\mathrm {CNT}} +Z_0 \tanh(\gamma l)} \right)^{-1}.
\label{eq:inputZ}
\end{equation}
$\gamma = \alpha+i\beta$ is the propagation constant with $\alpha$ and $\beta=2\pi\sqrt{\epsilon}f/c$ being the loss and phase constants respectively, $\epsilon$ the effective dielectric constant, $c$ the speed of light, $f$ the frequency, $Z_0$ the transmission line impedance, and the complex admittance $1/Z_{\mathrm {CNT}} = Y = \Gcnt + iB_{\mathrm{CNT}}$ with $\Gcnt$ conductance, and $B_{\mathrm{CNT}}$ susceptance of the CNT device.

The circuit is sensitive to changes in both the conductance and susceptance, as shown in Fig.~2a, where we plot a set of reflection spectra for different load admittances $Y$, with lengths chosen for a resonance at 3 GHz and a matched load of $4.8~\mu$S, assuming no losses. Varying $\Gcnt$ from $3~\mu\mathrm{S}$ to $4~\mu\mathrm{S}$ increases both the depth and width of the resonance, but not the resonance frequency. In contrast, a change in the susceptance $B$ mainly shifts the resonance, with a minor change in its depth. To illustrate the effects that losses have on this device, we can replace the circuit by an equivalent lossless one while introducing a phenomenological conductance $\GStub$ parallel to the CNT load (Fig.~1c) and thus defining a total effective load $\Geff=\Gcnt+\GStub$  seen by the microwaves. In Fig.~2b, we plot $\Gamma$ at the resonance as a function of $\Geff$. The reflected signal is the smallest close to the matched load offering maximal power transfer. The latter is more important for measuring small noise-power emitted from the device. Due to the additional loss conductance $\GStub$, the CNT load which matches the circuit is smaller than $\Gm$ and given by $\Gcnt^{\mathrm {Match}} = \Gm -\GStub$. In particular, if $\GStub > \Gm$ full matching is precluded and the reflection coefficient as a function of load becomes monotonic (see inset of Fig.~2b).

%%%%%%%%%%%%%%%%%%%%%%%%%%%%%%%%%%%%%%%%%%%%%%%%%%%%%%%%%%%%
%Figure3
%%%%%%%%%%%%%%%%%%%%%%%%%%%%%%%%%%%%%%%%%%%%%%%%%%%%%%%%%%%%
\begin{figure*}[t!]
 \centering
\includegraphics{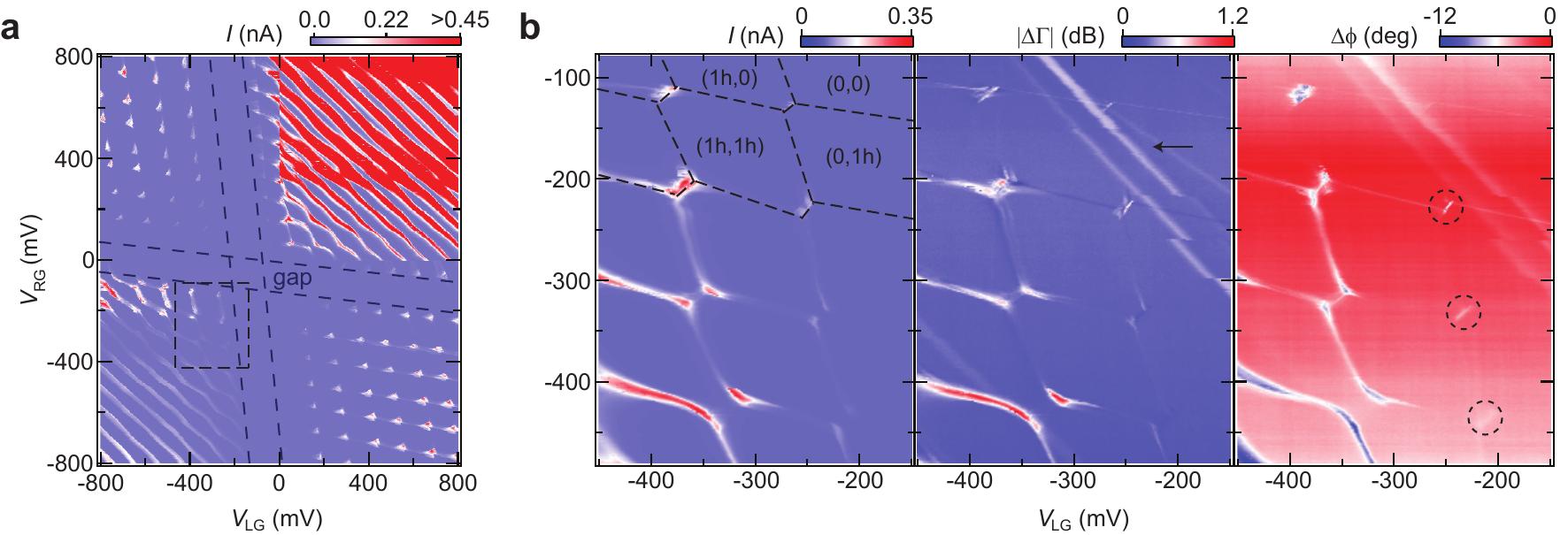} % for preprint
  \label{Figure3}
  \caption{\textbf{Charge stability and RF reflectometry of CNT double quantum dots.} 
	(\textbf{a}) DC measurements of the CNT device at $\Vsd=10~\mV$ and $\VMG=0$ show a semiconducting gap and the formation 
	of bipolar double dots around it when   $\VLG$ and $\VRG$ are swept. We observe the addition of electrons and holes in 
	each regime starting from complete depletion   in the bandgap. n-n double dots show relatively large conductance due 
	to n-doping of the contacts.
	(\textbf{b}) Simultaneous measurements of current, reflected amplitude and phase at the resonance frequency in the p-p 
	double dot region at $-10~\mV$ bias   indicated by a dashed box in panel (a). Co-tunneling lines are clearly visible in
	all the plots. Inter-dot coupling lines not present   in DC plot are visible in the amplitude and phase response due to 
	frequency shifts at gate voltages marked by dashed circles. The arrow points at spurious gate-tunable lines most likely 
	resulting from charge traps which do not necessarily contribute to the current but do change the susceptance. A probe power
	of -110~dBm is applied.}
\end{figure*}

The presented CNT device has three recessed local bottom gates with additional source and drain contacts elevated by $\sim 150$ nm (Fig.~1b). Bottom gates are separated by $\approx200$~nm from each other and from the source drain contacts producing a suspended length of the CNT of $\approx 800$~nm. The transmission lines are patterned on a Nb film and yield a resonance frequency $\fr/2\pi=2.9~\GHz$. All measurements are performed at the base temperature of the cryostat $\sim 20$~mK (Supplementary Note~3). The source is connected to the central conductor of the transmission line ($l$ branch) while the drain is connected to the ground plane. For the characterization of the devices, the CNT is tuned into the bandgap to present an infinite resistance and hence an open end. By measuring the reflection spectrum, it is possible to extract the relevant parameters ($l,d$ $\alpha$ and $\epsilon$). The extracted loss $\alpha= [0.0074,0.0082]~\mathrm{m}^{-1}$ for probe power in the range [-110,-140]~dBm corresponds to internal quality factors of 10,000 to 9000 for an equivalent half-wave resonator, showing that we are able to achieve low-loss microwave circuits in combination with CNT devices. Due to parasitic inductances from the fabricated contacts, the matching circuit has a lower effective $\Gm \approx 1.6~\mu$S compared to $\GStub\approx 3.2~\mu$S and operates therefore in the internal-loss-dominated regime. Measurements of the matching circuit response can be found in the Supplementary Note~4, as well as the measurement of an additional sample where full matching is demonstrated (Supplementary Note~5). \\
%%%%%%%%%%%%%%%%%%%%%%%%%%%%%%%%%%%%%%%%%%%%%%%%%%%%%%%%%%%%
%Figure4
%%%%%%%%%%%%%%%%%%%%%%%%%%%%%%%%%%%%%%%%%%%%%%%%%%%%%%%%%%%%
\begin{figure*}[t]
\includegraphics{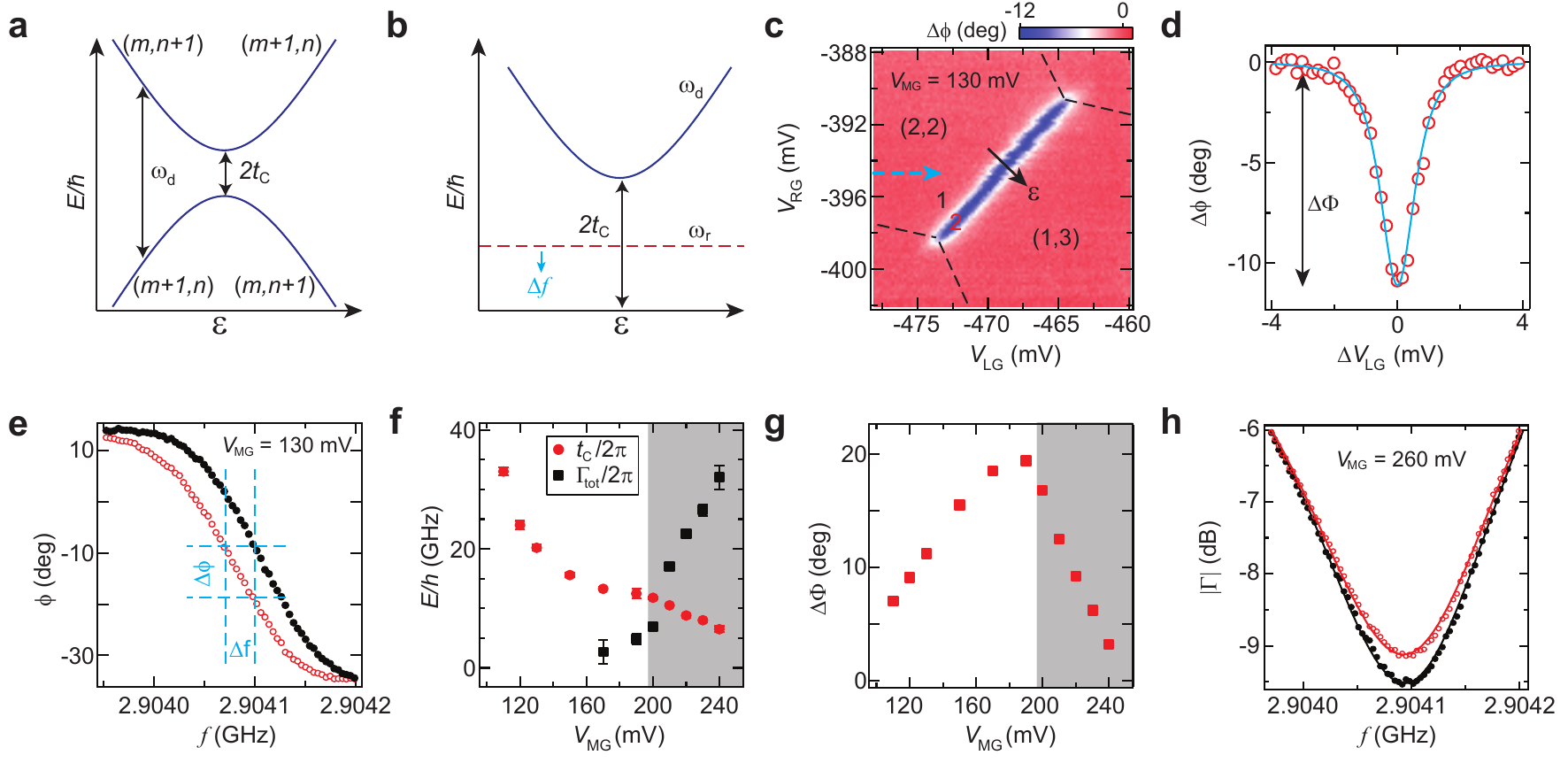} % for preprint
  \label{Figure4}
  \caption{\textbf{Reflectometry response of inter-dot coupling.} 
	(\textbf{a}) Energy levels of a double quantum dot with an
  inter-dot hybridization energy of $\tcoupl$, forming a two level system with energy $\fd$. 
	(\textbf{b}) The stub tuner
  resonance experiences a negative dispersive shift $\Delta f$ for $\fr<2\tcoupl$ due to its coupling to the two level system.
  (\textbf{c}) The phase response measured at the resonance frequency as a function of $\VLG$ and $\VMG$ yields a pronounced
  inter-dot transition due to hybridization of the charge states (2,2) and (1,3). 
	(\textbf{d}) A cut indicated by the dashed arrow in   (c) with a fit to eq.~2 to extract the inter-dot coupling energy $\tcoupl=20~\GHz$. 
	(\textbf{e}) The stub tuner phase response   at $\VMG=130~\mV$ around the dot degeneracy (open circles) and uncoupled dots 
	(solid circles) marked by points 2 and 1 in panel   (c) respectively. 
	(\textbf{f}) Extracted $\tcoupl$ and $\Gamma_{\mathrm {tot}}$ as a function of $\VMG$. Error bars represent uncertainties in the least square fitting. 
	(\textbf{g}) The maximum phase response $\Delta \Phi$ indicated in panel (d) is suppressed at larger $\VMG$ due to increasing $\Gamma_{\mathrm {tot}}$
  marked by the gray area.
	(\textbf{h}) The reflected power response at $\VMG = 260~\mV$ displays near-resonant absorption despite
  negligible dispersion. Solid lines are the fits to eq.~2. Symbols have the same meaning as in panel (e).}
\end{figure*}

\noindent \textbf{Locally tunability and RF reflectometry measurements of CNT double quantum dots.} Room temperature DC characterization during the device assembly allows us to perform several CNT mechanical transfer trials on the same device and choose nano\-tubes based on their gate dependence. In particular for semiconducting nanotubes, we can use the
bottom gates to locally shift the Fermi level above or below the valence and conduction bands of the nanotube. The latter allows for creating gate-defined confinement potentials along the nanotube and hence for tuning the location, size and number of quantum dots~\cite{Waissman2013}. With no RF power applied to the stub tuner and the middle gate $\VMG =0~$V, we first measure the charge stability diagram of the CNT device. A gate sweep using left ($\VLG$) and right gates ($\VRG$) at $\Vsd=10~\mV$ DC bias is shown in the Fig.~3a. The current response clearly displays the ambipolar behavior of quantum dots around a semiconducting gap of $\sim30~\meV$ with left and right gates tuning the CNT into n-n, n-p, p-n and p-p double dots. Here we have used a gate lever arm of $\approx 0.2$~meV/mV extracted from the Coulomb diamond measurements.  In addition, we observe the exact charge occupation of electrons and holes at corresponding gate voltages~\cite{Pei2012}. The high conductance in the n-n regime is possibly due to n-doping near the source/drain contacts. We also found p-doping for many samples for the same Pd contacts and do not exactly understand the nature of the observed contact doping.

We now perform simultaneous measurements of DC and RF reflectometry to obtain a collective response of the double quantum dots. $\Gamma$ is measured near the resonance frequency with a probe power chosen to be so low that no out-of-equilibrium charges are induced when applying microwave signal to the stub tuner. Figure~3b shows a qualitatively similar honeycomb charge stability diagram in current, amplitude and phase responses taken at $\Vsd = -10$ mV bias. We clearly observe co-tunneling lines~\cite{Frey2012}, long edges of the honeycomb, in all plots suggesting good sensitivity to impedance changes even in case of a circuit whose internal losses dominate the quality factor of the resonances. Here, we note an important distinction of the stub tuner. For half wave resonators, RF signal for co-tunneling lines strongly depend on strength of capacitive coupling to respective lead-dot transitions and their rates with respect to the resonance frequency~\cite{Frey2012b}. DC coupled stub tuner in contrast still responds through conductance changes that provide an external coupling by shunting microwaves via drain contact into the ground plane. This is further seen in Fig.~3b where larger current results in larger $\abs{\Delta\Gamma}$ and $\Delta \phi$ (see Supplementary Fig.~3a, 3b). In addition, we also observe hybridized double dots at degeneracy, the boundary of two honeycombs at the two smaller edges, marked by dashed circles in Fig.~3b, in the phase and amplitude plots. The signal results from the susceptance changes caused by dipole coupling of the hybridized charge states to the microwave resonance. The responses at different charge degeneracies are different due to the distinct dot coupling energies $\tcoupl$ which are affected by all gate voltages in our sample.

To illustrate the control over the confinement potential of the double quantum dots in this clean CNT device, $\VMG$ is used to tune the tunneling barrier between the two quantum dots. In the DC measurements, the strength of the tunneling coupling is visible as the separation between the charge triple points with the larger value corresponding to a stronger coupling or weaker barrier (Supplementary Fig.~5). For a quantitative analysis, the phase response of the stub tuner can be measured using a weak probe power (-130 dBm) near the hybridization of two charge states $(m,n+1)$ and $(m+1,n)$ as shown in Fig.~4a. We operate in the zero-bias regime, allowing the dots to stay in equilibrium and rule out any conductance changes which could affect the resonance response. Such a phase shift is shown in Fig.~4c close to (2,2) to (1,3) hole transition. We infer the frequency shifts $\Delta f$ from the phase variations which are almost linearly correlated near resonance [see Fig.~4e, error $< 10\%$ in our case]. Following a semiclassical model describing the coupling of a qubit with frequency  $\fd = \sqrt{\epsilon^2+4\tcoupl^2}$ to a resonator with frequency $\fr$ (Supplementary Note~7), the dispersion $\Delta f$ of the resonator is given by
\begin{equation}
2\pi\Delta f = -\textrm{Re}\left[\frac{(2\gzero\tcoupl/\fd)^2}{\Delta+i\Gamma_{\mathrm{tot}}} \langle\hat{\sigma}_{\mathrm Z}\rangle\right], \label{tc}
\end{equation}
where $\Gamma_{\mathrm{tot}} = \gamma/2+\Gamma_{\mathrm \phi}$ with $\gamma$ and $\Gamma_{\phi}$ the effective relaxation and dephasing rates of the hybridized double dot respectively, $\Delta = \fd-\fr$ and $\epsilon$ being the de\-tuning, $\gzero$ the zero temperature coupling strength with the resonator and $\langle \hat{\sigma}_{\mathrm Z}\rangle = \tanh(\hbar\fd/2\kBT)$ the polarization of the double dot transition at electronic temperature $T$. We use the equation to first extract $\gzero$ in a regime where $\tcoupl$ is large ($\VMG$ is small) so that $\fd \geq 2\tcoupl \gg \fr$, $\Gamma_{\mathrm{tot}}$. This yields a dependence of the frequency shift $\Delta f$ proportional to $\gzero/(\epsilon^2 + 4\tcoupl^2)^{3/2}$, now independent of $\Gamma_{\mathrm{tot}}$. A fit with this equation to the data at $\VMG = 130$~mV is shown in Fig.~4d, yielding $\gzero/2\pi=37~\MHz$. We find the same $\gzero$ for $\VMG =110$~mV supporting the assumption that  $\Gamma_{\mathrm{tot}}$ is relatively small in this regime. Fixing this $\gzero$ for phase responses at other $\VMG$, we plot the extracted $\tcoupl$ and $\Gamma_{\mathrm {tot}}$ in Fig.~4f and observe a reduction of $\tcoupl$ on increasing $\VMG$ reflecting a reduction in the tunnel coupling strength between the dots. The phase response starts to be suppressed for $\VMG$ larger than $200~\mV$ due to increasingly fast double dot relaxation, yielding $\Gamma_{\mathrm{tot}} > \tcoupl$. The inverse dependence of $\Gamma_{\mathrm \phi}$ on $\tcoupl$ has been seen in similar systems~\cite{Basset2013,Viennot2014} and could be due to the $1/f$ charge noise environment~\cite{Petersson2010} of the nanotube. The sign of the frequency shift always remains negative, further signifying that the resonator energy is always smaller than $2\tcoupl$. Increasing $\VMG$ to more than $250~\mV$, we do not notice any dispersion because of the large $\Gamma_{\mathrm {tot}}$. However, the hybridized dots still show a response similar to Fig.~2a, now  only in the reflectance amplitude. For $\VMG=260~\mV$, the stub tuner response for coupled and uncoupled dots regimes is presented in Fig.~4h. The fit to the resonance at the double dot degeneracy (open circles) shows a smaller depth and $\alpha= 0.0086$~m$^{-1}$ compared to the one in the uncoupled regime (solid circles) with $\alpha= 0.0082$~m$^{-1}$. This behavior is a result of an added loss channel i.e. absorption from the two-level hybridized dots when $2\tcoupl$ becomes comparable to $\fr$, which is also consistent with the change in the resonance depth due to the conductance increase for our device parameters~(Supplementary Fig.~3a,~6b). \\

\noindent \textbf{Discussion}

\noindent In summary, we have operated a RF superconducting impedance-matching circuit to measure CNT quantum dots in a hybrid device fabricated using a mechanical transfer of CNTs. The transfer employed here not only allows for a deterministic assembly and unity yield of complex RF devices, but also for the selection of CNTs with specific properties (metallic or semi-conducting), reuse of the same circuit with different tubes and incorporation of desirable contact materials. We demonstrate the ability to locally control the confinement potential along the length of the suspended nanotube and to form dots with a precise number of charges. The high symmetry in the ambipolar charge stability around the bandgap indicates low disorder in the CNT system.

Additionally, the matching circuit enables a comparatively simple extraction of the RF device impedance. We have shown that one can quantitatively deduce admittance changes in $\mu$S resolution by measuring the complex reflectance $\Gamma$. This sensitivity can be used to deduce basic parameters of a clean CNT double-dot operating
as a charge qubit, such as the interdot tunnel coupling strength and the relaxation rate. More importantly, conductances can be deduced using a simple analytic formula with  measurement BW reaching $(4/\pi) f_{\mathrm r}\zo\Gm$ estimated using eq.~1 near matching. We show an extraction of $G$ from the RF amplitude response for a similar device, now at full matching, in the Supplementary Note~5. We find Coulomb diamond plots quantitatively similar to its DC counterpart and demonstrate a BW up to 2~MHz and a reflectance down to -40~dB for device  $\Gcnt = 3.9~\mu$S. The reliable high-bandwidth extraction of $G$ at GHz frequencies holds promises for probing quantum charge-relaxation resistance, which can deviate from its usual DC counterpart described by the Landauer formula~\cite{Buttiker1993,Gabelli2006}. 

We demonstrate near matching with a substantial in and out microwave coupling in the measured device. For example, we see from Supplementary Fig.~3a that we achieve $|\Gamma|\approx -8$~dB at $G_{\mathrm{CNT}} \approx 0.4$~$\mu$S. This relates to a reflectance probability of 16~$\%$, hence 84~$\%$ is transmitted into the matching circuit and CNT device. Taking into account the internal loss described by $G_{\mathrm{Loss}}=3.2$~$\mu$S yields a substantial power transmission of $\sim 10$~$\%$ from a $2.5$M$\Omega$ device to a $50$\,$\Omega$ transmission line. This is beneficial for high throughput detection of emission noise from the quantum device defined in the CNT wire and shot noise measurements~\cite{Schoelkopf1998}. The presented ability to combine an intricate RF circuit with a pristine suspended CNT will invite novel studies on devices with engineered mechanical, electrical and photonic degrees of freedom. \\

\noindent \textbf{Methods}
\small

\noindent \textbf{Sample fabrication.} The device is patterned on a 150 nm thick Nb film, sputtered on an undoped Si substrate with 170 nm of $\mathrm{SiO}_2$, using photolithography and subsequent dry etching. The center conductor of the transmission line is $12~\mu$m wide, while gaps are $6~\mu$m wide, yielding a calculated $Z_0=49~\Omega$. The geometric lengths $l$ and $d$ of the stub tuner are 10.66 mm and 10.36 mm respectively. Spurious modes due to the T-junction are suppressed using on-chip wire bonds. The local bottom gates (Ti/Au of 5/35 nm thickness and 60 nm width) are defined by electron beam lithography and recessed in the $\mathrm{SiO}_2$ using dry anisotropic and isotropic etching with $\mathrm{CF}_4$ and Ar/$\mathrm{CHF}_3$, respectively (total recess depth $\approx$ 100 nm). A recess of depth $4~\mu$m is etched via dry etching with $\mathrm{SF}_6/\mathrm{O}_2$ around the Pd (100 nm) contacts to facilitate mechanical transfer~\cite{Muoth2013}. A modified micro-manipulator is used for CNT transfer at ambient conditions using optical microscopy. Contact resistances of CNT are generally found in excess of $100~\kohm$~\cite{Waissman2013}, which may be attributable to remaining oxide on the contacts and contact geometry. We obtained $20\%$ yield of transferring a single tube. More than one tube transferred at the same time due to multiple tubes on the same fork showed high conductance and could be removed by applying a large source drain bias.\\

\noindent \textbf{Data Analysis.} We extract the relevant parameters such as $l,d,\alpha$ of the device by fitting the resonance at zero conductance using the following equation
\begin{equation}
\Gamma = \frac{\mathrm{e}^{\mathrm{i}p} Z_{\mathrm {in}}-\zo}{\mathrm{e}^{\mathrm{i}p}Z_{\mathrm {in}}+\zo}+ \Gamma_0,
\end{equation}
where the phase factor $p$ accounts for impedance mismatches in the setup and $\Gamma_0$ for an offset. We could not exactly
determine the electronic temperature of the device and assume $T = 0~\K$ for all fits in Fig.~4d. For $\VMG < 160~\mV$,
extraction of relatively smaller $\Gamma_{\mathrm {tot}}$ is unreliable due to large detuning $\Delta$ from the stub tuner resonance.\\

%\noindent \bibliographystyle{naturemag}

%\noindent \bibliography{references_hybrid}
%%%%%%%%%%%%%%%%%%%%%%%%%%%%%%%%%%%%%%%%%%%%%%%%%%%%%%%%%%%%%%%%%%%%%%%%%%%%%%%%%%%%%%%%%%%%%%%%%%%%%%%%%%%%%%%%%%%%%%%%

%%%%%%%%%%%%%%%%%%%%%%%%%%%%%%%%%%%%%%%%%%%%%%%%%%%%%%%%%%%%%%%%%%%%%%%%%%%%%%%%%%%%%%%%%%%
\bigskip
\noindent \textbf{Acknowledgments} The authors thank J. Gramich for experimental assistance and acknowledge technical
support from the FIRST Center for Micro- and Nanoscience. We acknowledge financial support by the ERC project QUEST,
the EC project SE2ND, the NCCR QSIT and the Swiss National Science Foundation. \\

\noindent \textbf{Author Contributions} V. R. and G. P. H. contributed equally to the work. V. R. and G. P. H.
performed process development, sample fabrication, experiments and data analysis. M. J. and T. H. contributed to the
fabrication process development and measurements. M. M. and C. H. developed the transfer process and integrated the
carbon nano\-tubes. A. N. provided the theory support. G. P. H., A. W. and C. S. devised, initiated and supervised the work.
V. R. and G. P. H. wrote the manuscript with input from all authors.
%\newpage

\end{document}